\PassOptionsToPackage{unicode}{hyperref}
\PassOptionsToPackage{hyphens}{url}
\PassOptionsToPackage{dvipsnames,svgnames,x11names}{xcolor}
\documentclass[
  12pt]{article}

\usepackage{amsmath,amssymb}
\usepackage{iftex}
\ifPDFTeX
  \usepackage[T1]{fontenc}
  \usepackage[utf8]{inputenc}
  \usepackage{textcomp} 
\else 
  \usepackage{unicode-math}
  \defaultfontfeatures{Scale=MatchLowercase}
  \defaultfontfeatures[\rmfamily]{Ligatures=TeX,Scale=1}
\fi
\usepackage{lmodern}
\ifPDFTeX\else  
\fi
\IfFileExists{upquote.sty}{\usepackage{upquote}}{}
\IfFileExists{microtype.sty}{
  \usepackage[]{microtype}
  \UseMicrotypeSet[protrusion]{basicmath} 
}{}
\makeatletter
\@ifundefined{KOMAClassName}{
  \IfFileExists{parskip.sty}{%
    \usepackage{parskip}
  }{
    \setlength{\parindent}{0pt}
    \setlength{\parskip}{6pt plus 2pt minus 1pt}}
}{
  \KOMAoptions{parskip=half}}
\makeatother
\usepackage{xcolor}
\makeatletter
\ifx\paragraph\undefined\else
  \let\oldparagraph\paragraph
  \renewcommand{\paragraph}{
    \@ifstar
      \xxxParagraphStar
      \xxxParagraphNoStar
  }
  \newcommand{\xxxParagraphStar}[1]{\oldparagraph*{#1}\mbox{}}
  \newcommand{\xxxParagraphNoStar}[1]{\oldparagraph{#1}\mbox{}}
\fi
\ifx\subparagraph\undefined\else
  \let\oldsubparagraph\subparagraph
  \renewcommand{\subparagraph}{
    \@ifstar
      \xxxSubParagraphStar
      \xxxSubParagraphNoStar
  }
  \newcommand{\xxxSubParagraphStar}[1]{\oldsubparagraph*{#1}\mbox{}}
  \newcommand{\xxxSubParagraphNoStar}[1]{\oldsubparagraph{#1}\mbox{}}
\fi
\makeatother

\usepackage{longtable,booktabs,array}
\usepackage{calc} 
\usepackage{etoolbox}
\makeatletter
\patchcmd\longtable{\par}{\if@noskipsec\mbox{}\fi\par}{}{}
\makeatother
\IfFileExists{footnotehyper.sty}{\usepackage{footnotehyper}}{\usepackage{footnote}}
\makesavenoteenv{longtable}
\usepackage{graphicx}
\makeatletter
\def\maxwidth{\ifdim\Gin@nat@width>\linewidth\linewidth\else\Gin@nat@width\fi}
\def\maxheight{\ifdim\Gin@nat@height>\textheight\textheight\else\Gin@nat@height\fi}
\makeatother
\setkeys{Gin}{width=\maxwidth,height=\maxheight,keepaspectratio}
\makeatletter
\def\fps@figure{htbp}
\makeatother

\makeatletter
\@ifpackageloaded{caption}{}{\usepackage{caption}}
\AtBeginDocument{%
\ifdefined\contentsname
  \renewcommand*\contentsname{Table of contents}
\else
  \newcommand\contentsname{Table of contents}
\fi
\ifdefined\listfigurename
  \renewcommand*\listfigurename{List of Figures}
\else
  \newcommand\listfigurename{List of Figures}
\fi
\ifdefined\listtablename
  \renewcommand*\listtablename{List of Tables}
\else
  \newcommand\listtablename{List of Tables}
\fi
\ifdefined\figurename
  \renewcommand*\figurename{Figure}
\else
  \newcommand\figurename{Figure}
\fi
\ifdefined\tablename
  \renewcommand*\tablename{Table}
\else
  \newcommand\tablename{Table}
\fi
}
\@ifpackageloaded{float}{}{\usepackage{float}}
\floatstyle{ruled}
\@ifundefined{c@chapter}{\newfloat{codelisting}{h}{lop}}{\newfloat{codelisting}{h}{lop}[chapter]}
\floatname{codelisting}{Listing}

\makeatother
\makeatletter
\@ifpackageloaded{caption}{}{\usepackage{caption}}
\@ifpackageloaded{subcaption}{}{\usepackage{subcaption}}
\makeatother

\ifLuaTeX
  \usepackage{selnolig}  
\fi
\usepackage[]{natbib}
\usepackage{bookmark}

\IfFileExists{xurl.sty}{\usepackage{xurl}}{} 
\hypersetup{
  pdftitle={Expected Shortfall Model Averaging},
  pdfauthor={Jianming Wu; Xinyu Zhang; Jie Zeng},
  pdfkeywords={ Asymptotic optimality; Model averaging;  Expected Shortfall; Value-at-Risk},
  colorlinks=true,
  linkcolor={blue},
  filecolor={Maroon},
  citecolor={Blue},
  urlcolor={Blue},
  pdfcreator={LaTeX via pandoc}}

\usepackage{endnotes}
\usepackage{bm}
\usepackage{threeparttable}
\usepackage{algorithm}
\usepackage{algorithmic}

\newtheorem{thm}{\underline{\bf Theorem}}

\newtheorem{remark}{\underline{\bf Remark}}

\allowdisplaybreaks

\newcommand{\calD}{{\cal D}}

\def\wh{\widehat}
\def\wt{\widetilde}

\def\mR{\mathbb{R}}

\def\argmin{\mbox{argmin}}

\def\VaR{{\mbox{VaR}}}
\def\ES{{\mbox{ES}}}

\def\0{{\bm 0}}

\def\mE{\mathbb{E}}

\def\V{{\bm V}}

\def\u{{\bm u}}

\def\W{{\bm W}}

\def\w{{\bm w}}
\def\X{{\bm{X}}}

\def\Z{{\bm Z}}

\def\bnu{{\bm \nu}}

\def\bq{\begin{equation}}
	\def\eq{\end{equation}}

\def\wh{\widehat}
\def\wt{\widetilde}

\def\log{\hbox{log}}

\def\squarebox#1{\hbox to #1{\hfill\vbox to #1{\vfill}}}
\def\btheta{{\boldsymbol \theta}}
\def\bTheta{\bm{\Theta}}

\def\mH{\mathcal{H}}

\def\wh{\widehat}
\def\wt{\widetilde}

\def\log{\hbox{log}}

\def\boxit#1{\vbox{\hrule\hbox{\vrule\kern6pt\vbox{\kern6pt#1\kern6pt}\kern6pt\vrule}\hrule}}

\newcommand{\anon}{1}

\begin{document}

\def\spacingset#1{\renewcommand{\baselinestretch}%
{#1}\small\normalsize} \spacingset{1}


\if1\anon
{
  \title{\bf Expected Shortfall Model Averaging}
  \author{Jianming Wu$^{1}$, Xinyu Zhang$^{2,1}\footnote{Corresponding author: Xinyu Zhang, State Key Laboratory of Mathematical Sciences, Academy of Mathematics and Systems Science, Chinese Academy of Sciences, Beijing, China. Email: xinyu@amss.ac.cn.}$, Jie Zeng$^{3}$\\
  	$^{1}$School of Management, University of Science and Technology\\ of China\\
  	$^{2}$State Key Laboratory of Mathematical Sciences, Academy of\\ Mathematics and Systems Science, Chinese Academy of Sciences\\
  	$^{3}$School of Mathematics and Statistics, Hefei Normal University}
  \maketitle
} \fi

\if0\anon
{
  \bigskip
  \bigskip
  \bigskip
  \begin{center}
    {\LARGE\bf Title}
\end{center}
  \medskip
} \fi

\bigskip
\begin{abstract}
Expected shortfall (ES) is widely used to measure tail risk in finance and economics, but its prediction is challenging due to non-elicitability and model uncertainty. This paper proposes a two-stage cross-validation model averaging method for ES forecasting. In the first stage, conditional value-at-risk is estimated using quantile model averaging. In the second stage, a transformed response is constructed and mean squared error–based model averaging is applied to estimate ES.  We establish theoretical properties of the proposed method under both correct specification and model misspecification, showing consistency of the estimators and asymptotic optimality of the forecasting risk. Simulation studies and empirical applications to U.S. stock return and macroeconomic GDP growth data show that the proposed approach provides accurate and stable ES forecasts and is computationally efficient.
\end{abstract}

\noindent%
{\it Keywords:} Asymptotic optimality;  Forecast combination; Tail risk; Value-at-Risk
\vfill

\newpage
\spacingset{1.8} 

\section{Introduction}\label{chap:introduction}
\setcounter{equation}{0}
\baselineskip=24pt
Risk management in finance and economics has attracted increasing attention from policymakers. Accurate risk forecasting plays an important role in subsequent decision-making processes, such as policy design and portfolio optimization. Value-at-Risk (VaR), also known as a quantile-based risk measure, has been widely used due to its simplicity and ease of interpretation. However, as pointed out by many researchers, VaR suffers from several drawbacks in practice, including the lack of coherence. In recent years, Expected Shortfall (ES) has received growing interest in both academia and practice. For example, Basel III has adopted ES as a regulatory risk measure in financial reporting to better reflect downside risk. 

Specifically, let $Y$ be a real-valued random variable representing asset returns or gross domestic product (GDP) growth. Suppose $Y$ has a finite first absolute moment, that is, $\mathbb{E}|Y| < \infty$. Let $F_Y$ denote the cumulative distribution function of $Y$. For any $\tau \in (0,1)$, the VaR and ES at level $\tau$ are defined as
\begin{align*}
	\VaR_\tau(Y) &= F_Y^{-1}(\tau)
	= \inf\{y \in \mathbb{R} : F_Y(y) \ge \tau\}
\end{align*}
and
\begin{align*}
	\ES_\tau(Y) &= \mathbb{E}[Y | Y \le \VaR_\tau(Y)],
\end{align*}
respectively. If $F_Y$ is continuous, the ES can equivalently be expressed as
\begin{align*}
	\ES_\tau(Y) = \frac{1}{\tau} \int_0^\tau \VaR_u(Y)\mathrm{d}u.
\end{align*}
As the conditional mean below the $\tau$-quantile, ES provides additional information on the tail behavior beyond VaR.

Despite its advantages, a major challenge in ES estimation is that ES is not elicitable on its own. A risk measure (or more generally, a statistical functional) is said to be \textit{elicitable} if it can be characterized as the minimizer of the expected value of a suitable loss function \citep{Gneiting2011making}. This property is crucial for regression modeling and forecast evaluation. Recently, \citet{Fissler2016Higher} showed that VaR and ES are jointly elicitable and proposed a class of joint loss functions. This result has stimulated substantial research on joint VaR–ES regression models. For instance, \citet{Dimitriadis2019joint} proposed an M-estimation approach under independence assumptions, while \citet{Patton2019Dynamic} developed a dynamic joint model for time-series data using M-estimation.

One practical difficulty of M-estimation based on joint loss functions is that the resulting objective functions are often non-differentiable and non-convex, which complicates computation. From a different perspective, \citet{barendseefficiently} proposed a two-stage estimation procedure based on Neyman orthogonality to avoid such optimization problems. Neyman-orthogonal scores reduce the impact of nuisance parameter estimation on the target parameter. This framework was further extended to heavy-tailed data by \citet{He2023}. In a nonparametric setting, \citet{Yu2025estimation} proposed a two-step ES estimator in a reproducing kernel Hilbert space. 

Another important challenge in ES regression is model uncertainty, which is common in financial and economic applications.  Model uncertainty arises when there are many competing candidate models, and it is unclear which one should be used for prediction. A traditional approach to addressing model uncertainty is model selection, such as information criteria \citep{Akaike1973maximum} or regularization methods \citep{Tibshirani1996regression}. Within the two-stage ES regression framework, \citet{Zhang2025} proposed an $\mathcal{L}_1$-type regularization method for feature selection in high-dimensional linear models. However, model selection procedures are known to be unstable \citep{Yuan2005combining}. Model averaging provides an alternative approach by combining multiple candidate models through weighted averaging, thereby reducing the risk of selecting a poorly performing model \citep{Hsiao2014there,Racine2023Optimal}.  Depending on the estimation target, existing model averaging methods have been developed for a variety of problems, including mean estimation \citep{Hansen2007,zhang2022model}, quantile estimation \citep{Lu2015,Tu2025Quantile}, and distribution estimation \citep{Nelson2021reducing}. More recently, \citet{Jiao2025Optimal} developed a one-step jackknife model averaging method for joint VaR–ES forecasting based on the joint loss function in \citet{Fissler2016Higher}. However, this joint model averaging procedure can be computationally intensive due to the complex structure of the loss function.  Moreover, the framework implicitly focuses on cases with negative ES values, which may limit its applicability in situations where ES can be nonnegative, such as periods with consistently strong positive returns.


To address these issues, we  propose a two-stage cross-validation model averaging approach for expected shortfall forecasting (ESMA). In the first stage, we apply existing quantile model averaging methods to estimate a weighted VaR. In the second stage, conditional on the first-stage VaR estimator, we construct a cross-validation criterion based on mean squared error to obtain optimal weights for ES estimation.

The contributions of this paper are threefold. First, we propose a two-stage model averaging procedure based on $K$-fold cross-validation. The method is computationally efficient, as it relies on standard quantile and mean estimation, which can be implemented using linear and quadratic programming. Second, we establish theoretical properties for the proposed method.  The analysis is nontrivial because the second-stage ES regression is built on a pseudo response generated by the first-stage model averaging estimator, so the second-stage target and risk both depend on first-stage estimation error. This creates a  two-layer problem and requires us to control the propagation of first-stage error under both correct specification and global misspecification. Allowing for misspecified candidate models, we derive the uniform convergence rate for the second-stage coefficient estimators. When the set of candidate models contains the true model, we show that the ES model averaging estimator is consistent and that the sum of the weights converges to one. When all candidate models are misspecified, we prove that the proposed method is asymptotically optimal. To the best of our knowledge, this is the first two-stage model averaging approach. Third, we evaluate the proposed method through simulation studies and empirical applications. The results show that the ESMA method delivers more accurate forecasts than existing joint model averaging and two-stage model selection approaches.

The remainder of this paper is organized as follows. Section \ref{chap:method} introduces the two-stage model averaging framework. Section \ref{chap:theory} presents the theoretical results. Sections \ref{chap:sim} and \ref{chap:empirical} report simulation and empirical studies, respectively. Section \ref{chap:conclusion} concludes. All proofs and additional experimental results are provided in the Online Supplement.


\section{Methodology}\label{chap:method}
Consider a sequence of observations $\calD_n=\{(Y_i,\X_i)\}_{i=1}^n$, where $\X_i$ is the predictor vector and may be infinite-dimensional. The conditional expected shortfall is defined as
\begin{align}
	\ES_\tau(Y_i|\X_i)=\mathbb{E}\{Y_i| Y_i\le \VaR_\tau(Y_i|\X_i),\X_i\}, \label{eq.condition_es}
\end{align}
where the $\tau$-level conditional quantile is $\VaR_\tau(Y_i|\X_i)=F_{Y_i|\X_i}^{-1}(\tau)$, and $F_{Y_i|\X_i}$ denotes the conditional distribution function given $\X_i$. Given a new observation $\X_0$ with unknown response $Y_0$, our goal is to predict the conditional expected shortfall of $Y_0$, i.e., $\ES_\tau(Y_0|\X_0)$.

In this paper, we consider the following data generating process (DGP):
\begin{align}
	Y_i=\sum_{j=1}^\infty\beta_jX_{ij}+\left(\sum_{j=1}^\infty\alpha_jX_{ij}\right)\eta_i,\ i=1,2,\ldots,n,\label{eq.dpg}
\end{align}
where $\eta_i$ are independent and identically distributed (i.i.d.) random innovations, independent of the covariates $X_{ij}$. This DGP allows for heteroskedasticity through covariate-dependent scale effects.  To avoid the quantile crossing problem \citep[see, e.g.,][]{he1997quantile}, we assume that the scale components $\sum_{j=1}^\infty\alpha_jX_{ij}$ are positive almost surely. By definition \eqref{eq.condition_es}, the $\tau$-level VaR and ES take the forms
\begin{align*}
	\VaR_\tau(Y_i|\X_i)=&\sum_{j=1}^\infty\{\beta_j+\alpha_j\VaR_\tau(\eta_i)\}X_{ij}\equiv\sum_{j=1}^\infty\nu_jX_{ij}
\end{align*}
and
\begin{align*}
	\ES_\tau(Y_i|\X_i)=&\sum_{j=1}^\infty\{\beta_j+\alpha_j\ES_\tau(\eta_i)\}X_{ij}\equiv\sum_{j=1}^\infty\theta_jX_{ij},
\end{align*}
where $\X_i=(X_{i1},X_{i2},\ldots)^\top$ and $X_{i1}=1$. In matrix form, we write
\begin{align*}
	\VaR_\tau(Y_i|\X_i)=\X_i^\top\boldsymbol{\nu}_0\text{ and }
	\ES_\tau(Y_i|\X_i)=\X_i^\top\boldsymbol{\theta}_0,
\end{align*}
where $\boldsymbol{\nu}_0=(\nu_1,\nu_2,\ldots)^\top$ and $\boldsymbol{\theta}_0=(\theta_1,\theta_2,\ldots)^\top$ denote the true coefficient vectors. Since the dimension in \eqref{eq.dpg} may be infinite, this framework includes nonparametric regression as a special case when $\X_i$ contains basis expansions.

In practice, the full covariate vector $\X_i$ is unobservable. Instead, we only observe a subset of predictors, possibly together with irrelevant variables. This misspecification and uncertainty motivate a model averaging procedure to combine all possible models. Our model averaging procedure consists of two stages. In the first step, we consider $M_1$ candidate models. The $m$-th model uses  a $p_m$-dimensional vector $\X_{(m),i}$. For $i=0,1,\ldots,n$, let $\wt{\X}_i$ denote the $\tilde{p}$-dimensional vector that collects all predictors appearing in the $M_1$ candidate models. For each $1\le m\le M_1$, let ${\bm P}_{(m)}^Q$ be a $p_m \times \widetilde{p}$ selection matrix such that $\X_{(m),i} = {\bm P}_{(m)}^Q \widetilde{\X}_i$. For each candidate model, we estimate $\widehat{\boldsymbol{\nu}}_{(m)}$ by standard quantile regression:
\begin{align*}
	\widehat{\boldsymbol{\nu}}_{(m)} = \argmin_{\boldsymbol{\nu}_{(m)}}\sum_{i=1}^n\rho_\tau(Y_i-\X_{(m),i}^\top\boldsymbol{\nu}_{(m)}),
\end{align*}
where $\rho_\tau(u)=(\tau-\mathbb{I}(u<0))u$ is the check loss. We estimate the model averaging weights $\widehat{\w}_Q= (\wh{w}_1^Q,\ldots,\wh{w}_{M_1}^Q)^\top$ via $K$-fold cross-validation following \citep{Gu2025Model}. The resulting model averaging coefficient is $\widehat{\V}(\widehat{\w}_Q)= \sum_{m=1}^{M_1}\wh{w}_m^Q({\bm P}_{(m)}^{Q})^\top\wh{\bnu}_{(m)}$, and the corresponding VaR prediction is $\wt{\X}_0^\top\widehat{\V}(\widehat{\w}_Q)$.

Define the pseudo response with any $\tilde{p}$-dimensional $\V$ as
\begin{align}
	\widetilde{Y}_i(\V)=\tau^{-1}(Y_i-\wt{\X}_i^\top\V)\mathbb{I}(Y_i\le\wt{\X}_i^\top\V)+\wt{\X}_i^\top\V.
\end{align}
Let $\V_0$ denote the true coefficient vector associated with $\wt{\X}_i$. If $\wt{\X}_i$ includes all informative predictors in $\X_i$, then we have $\mE\{\widetilde{Y}_i(\V_0)|\X_i\}=\ES_\tau(Y_i|\X_i)$ almost surely. This conditional mean representation provides an alternative to joint estimation of VaR and ES and leads to a computationally convenient two-stage procedure. In the second stage, we use $\widetilde{Y}_i(\widehat{\V}(\widehat{\w}_Q))$ as the response and mean squared error (MSE) as the loss function. 

We consider $M_2$ candidate models in the second stage, which may differ from those in the first stage.  Both $M_1$ and $M_2$ are allowed to diverge with the sample size. The $l$-th model uses a $q_l$-dimensional predictor vector $\Z_{(l),i}$. Let $\widetilde{\Z}_i$ collect all predictors appearing in the $M_2$ models. For each $l$, let ${\bm P}_{(l)}^E$ denote the selection matrix such that $\Z_{(l),i}={\bm P}_{(l)}^E\widetilde{\Z}_i$. For each candidate model, we estimate $\widehat{\boldsymbol{\theta}}_{(l)}$ by minimizing the loss $\sum_{i=1}^n(\widetilde{Y}_i(\widehat{\V}(\widehat{\w}_Q))-\Z_{(l),i}^\top\boldsymbol{\theta}_{(l)})^2$. Let $\wh{\bTheta}_{(l)}=({\bm P}_{(l)}^E)^\top\widehat{\boldsymbol{\theta}}_{(l)}$ denote the coefficient of $\wt{\Z}_i$ in the $l$-th model and $\wh{\bTheta}(\w)= \sum_{l=1}^{M_2}w_l\wh{\bTheta}_{(l)}$ denote the model averaging coefficient. The weights $\widehat{\w}_{ES}$ are estimated by $K$-fold cross-validation. Specifically, we first divide the $n$ samples into $K\ge2$ groups $\{I_k\}_{k=1}^K$, where $\cup_{1\le k\le K} I_k=\{1,2,\ldots,n\}$ and $I_j\cap I_k=\emptyset$ for any $j\ne k$. Without loss of generality, we assume each group contains $J=n/K$ samples. For the $k$-th group, we use the remaining  samples in $\cup_{j\ne k}I_j$ to estimate $\wh{\btheta}_{(l)}^{[-k]}$, that is, $\wh{\btheta}_{(l)}^{[-k]}=\argmin_{\btheta_{(l)}}\sum_{i\in \cup_{j\ne k}I_j}(\widetilde{Y}_i(\widehat{\V}(\widehat{\w}_Q))-\Z_{(l),i}^\top\boldsymbol{\theta}_{(l)})^2$.  Let $\wh{\bTheta}_{(l)}^{[-k]}=({\bm P}_{(l)}^E)^\top\widehat{\boldsymbol{\theta}}_{(l)}^{[-k]}$ and $\wh{\bTheta}^{[-k]}(\w)=\sum_{l=1}^{M_2}w_l\wh{\bTheta}_{(l)}^{[-k]}$. The cross-validation criterion is given by
\begin{align*}
	CV_{ES}(\w)=\frac{1}{n}\sum_{k=1}^K\sum_{i\in I_k}\left(\widetilde{Y}_i(\widehat{\V}(\widehat{\w}_Q))-\wt{\Z}_i^\top\wh{\bTheta}^{[-k]}(\w)\right)^2.
\end{align*}

The optimal weights are defined as 
\begin{align*}
	\wh{\w}_{ES}=\argmin_{\w\in \mH_{M_2}}CV_{ES}(\w),
\end{align*}
where $\mH_{M_2}=\{\w\in [0,1]^{M_2}: \|\w\|_1=1\}$. The final expected shortfall prediction is $\wt{\Z}_0^\top\widehat{\bTheta}(\widehat{\w}_{ES})$.\endnote{
	For applications requiring finite-sample coherence at a given forecast
	point, each second-stage coefficient estimator, including its
	fold-specific version, can be replaced by a constrained estimator
	$\widehat{\btheta}_{(l)}^{\,c}$ satisfying
	$\Z_{(l),0}^\top\widehat{\btheta}_{(l)}^{\,c}
	\le
	\widetilde{\X}_0^\top\widehat{\V}(\widehat{\w}_Q)$.
	Writing
	$\widehat{\bTheta}^{\,c}(\w)
	=
	\sum_{l=1}^{M_2}w_l
	({\bm P}_{(l)}^E)^\top
	\widehat{\btheta}_{(l)}^{\,c}$,
	the same ordering restriction can also be imposed directly on the
	averaged forecast during weight optimization:
	$\widetilde{\Z}_0^\top\widehat{\bTheta}^{\,c}(\w)
	\le
	\widetilde{\X}_0^\top\widehat{\V}(\widehat{\w}_Q)$.
	These linear restrictions yield computationally tractable constrained
	least-squares problems. The analysis in
	this paper concerns the unconstrained estimator.}

\begin{remark}
	In the first stage, $\widehat{\V}(\widehat{\w}_Q)$ is estimated once
	using the full sample. Consequently, although the observations in
	$I_k$ are excluded when estimating the fold-specific second-stage
	coefficients, they still enter the held-out pseudo responses through
	the first-stage estimator. The resulting second-stage criterion is
	therefore not based on fully nested cross-validation.
	
	Proposition S.1 in the Online Supplement shows that the dependence
	induced by this reuse of the data is asymptotically negligible. Let
	$\V_\circ$ denote the nonrandom target associated with
	$\widehat{\V}(\widehat{\w}_Q)$, and let $CV_{ES}^{\circ}(\w)$ denote
	the oracle criterion obtained by replacing the estimated first-stage
	coefficient with $\V_\circ$. Since $\V_\circ$ is nonrandom, the oracle
	criterion preserves the foldwise separation between estimation and
	validation. Under the stated rate and moment conditions,
	\[
	\sup_{\w\in\mH_{M_2}}
	\left|
	\frac{CV_{ES}(\w)}
	{CV_{ES}^{\circ}(\w)}
	-1
	\right|
	=o_p(1).
	\]
	This result holds both when the first-stage candidate set contains a
	correctly specified model and when all first-stage candidate models
	are misspecified. Consequently, the optimal weight $\wh{\w}_{ES}$ asymptotically minimizes the oracle criterion.
\end{remark}

\begin{remark}
	To obtain robustness under sub-Gaussian or heavy-tailed data, \citet{He2023} proposed a robust two-step method that employs an adaptive Huber loss in the second stage. In our framework, the squared loss can similarly be replaced by the Huber loss. However, selecting the Huber tuning parameter under model misspecification poses theoretical challenges. We leave this issue for future research.
\end{remark}

\section{Theoretical Results}
\label{chap:theory}
We use $\|\cdot\|$ to denote the $\mathcal{L}_2$-norm.  Let $\lambda_{\min}(A)$ and $\lambda_{\max}(A)$ denote the minimum and maximum eigenvalues of matrix $A$, respectively. Without loss of generality, we assume that all random predictors have zero means. Let $\boldsymbol{\nu}_{(m)}^*=\argmin_{\boldsymbol{\nu}_{(m)}}\mE\{\rho_\tau(Y_i-\X_{(m),i}^\top\bnu_{(m)})\}$ denote the pseudo-true parameter under the $m$-th candidate model. Let $\bar{q}=\max_{1\le l\le M_2}q_l$ denote the maximum dimension in the second stage. For identification, we assume $\bar{q}\le n$.

\noindent\textbf{Assumption 1} For $1\le l\le M_2$, the covariate vector $\Z_{(l)}\in\mR ^{q_l}$ is sub-Gaussian. That is, there exists a constant $c_1\ge1$ such that $\Pr(|\u^\top\W_{Z,(l)}|\ge c_1t)\le 2\exp(-t^2/2)$ for all $t\ge 0$ and $\u\in \mathbb{S}^{q_l-1}$, where $\W_{Z,(l)}=\Sigma_{Z,(l)}^{-1/2}\Z_{(l)}$, $\Sigma_{Z,(l)}=\mE(\Z_{(l)}\Z_{(l)}^\top)$ and $\mathbb{S}^{q_l-1}$ denotes the $(q_l-1)$-dimensional unit sphere. 

\noindent\textbf{Assumption 2} The conditional density $f_{\varepsilon|\X}$ of $\varepsilon=Y-\X^\top\bnu_0$ given $\X$ exists and is continuous on its support. Moreover, the conditional distribution function $F_{\varepsilon|\X}$ is continuously differentiable and satisfies $|F_{\varepsilon|\X}(t)-F_{\varepsilon|\X}(0)|\le \bar{f}|t|$ for some constant $\bar{f}$ and all $t\in\mR$.

\noindent\textbf{Assumption 3} $\max_{1\le m\le M_1}\|\wh{\bnu}_{(m)}-\bnu_{(m)}^*\|=O_p(r^Q)$. 

Assumption 1 controls the tail behavior of the second-stage covariates and ensures uniform concentration of empirical quantities around their population counterparts. Similar conditions are commonly imposed in high-dimensional regression (e.g., Condition 2 of \citet{He2023}). Assumption 2 provides mild smoothness conditions on the conditional error distribution and allows for conditional heteroskedasticity. Assumption 3 is a high-level condition requiring uniform convergence of the first-stage quantile estimators with rate $r^Q$. Under regularity conditions, \citet{Lu2015} show that $r^Q=n^{-1/2}\bar{p}^{1/2}(\log n)^{1/2}$, where $\bar{p}=\max_{1\le m\le M_1}p_m$.

We first study the theoretical properties when there exist correct models. Following \citet{Zhang2020parsimonious}, a \textit{true model} is defined as a model that includes all and only regressors with nonzero coefficients. A \textit{correct model} is one that contains all regressors in the true model, possibly along with additional irrelevant variables. Let $\mathcal{D}^Q$ and $\calD^E$ denote the sets of correct models in the first and second stages, respectively. For each $m\in\calD^Q$, the covariates $\X_{(m),i}$ include the true regressors and possibly some irrelevant ones. As a result, the pooled covariate vector $\wt{\X}_i$ contains the true predictors and additional noise variables. Existing results in quantile regression \citep[e.g.][]{He2023} imply $\|\wh{\bnu}_{(m)}-\bnu_{(m)}^*\|=O_p(n^{-1/2}\bar{p}^{1/2})$, where $\bnu_{(m)}^*$ consists of the true coefficients padded with zeros. Let $\Sigma_X=\mE(\wt{\X}\wt{\X}^\top)$.

\noindent\textbf{Assumption 4}  $\|\wh{\V}(\wh{\w}_Q)-\V_0\|_{\Sigma_X}=O_p(r_0)$.

Assumption 4 requires the first-stage model averaging estimator to converge to the true coefficient vector at rate $r_0$, which depends on the sample size $n$, the number of candidate models $M_1$, and the maximal model dimension $\tilde{p}$. Under regularity conditions, \citet{Gu2025Model} show that $r_0=\bar{c}_X^{1/2}((r^Q)^{1/2}\tilde{p}^{1/2}+n^{-1/4}M_1^{1/4})$ with $\bar{c}_X=\lambda_{\max}(\Sigma_X)$. Define the pseudo-true transformed response
\begin{align}
	\widetilde{Y}_i^*=&\tau^{-1}(Y_i-\wt{\X}_i^\top\V_0)\mathbb{I}(Y_i\le\wt{\X}_i^\top\V_0)+\wt{\X}_i^\top\V_0\nonumber\\
	=&\tau^{-1}(Y_i-\X_i^\top\bnu_0)\mathbb{I}(Y_i\le\X_i^\top\bnu_0)+\X_i^\top\bnu_0.
\end{align}
By \eqref{eq.condition_es} and the DGP \eqref{eq.dpg}, we have $\mE[\widetilde{Y}_i^*| \wt{\X}_i]=\ES_\tau(Y_i| \X_i)$ almost surely. For the second stage, define the pseudo-true ES parameter
\begin{align*}
	\btheta_{(l)}^*=\argmin_{\btheta_{(l)}}\mE(\wt{Y}_i^*-\Z_{(l),i}^\top\btheta_{(l)})^2.
\end{align*}
If the $(M_2+1)$-th model is correctly specified, then $\Z_{(M_2+1),i}^\top\btheta_{(M_2+1)}^*=\Z_i^\top\btheta_0$ almost surely.

\noindent\textbf{Assumption 5} The covariate vector $\wt{\X}$ is sub-Gaussian. That is,  $\Pr(|\u^\top\W_{X}|\ge c_1t)\le 2\exp(-t^2/2)$ for $\u\in\mathbb{S}^{\tilde{p}-1}$, where  $\W_X=\Sigma_X^{-1/2}\wt{\X}$. In addition, $\sup_{1\le l\le M_2}\mE[(\epsilon_{(l),i}^{*})^4|\wt{\Z}_i]\le \bar{\sigma}^4<\infty$ almost surely over $\wt{\Z}_i$, where $\epsilon_{(l),i}^*=\wt{Y}_i^*-\Z_{(l),i}^\top\btheta_{(l)}^*$ and $\bar{\sigma}$ is some positive constant.

Assumption 5 imposes a sub-Gaussian condition on the first-stage covariates and ensures uniformly bounded fourth moments for the pseudo-innovations in the second stage. Define $\kappa_{Z,k}=\sup_{1\le l\le M_2}\sup_{\u\in \mathbb{S}^{q_l-1}}\mE|\u^\top\W_{Z,(l)}|^k$ and $\kappa_{X,k}=\sup_{\u\in \mathbb{S}^{\tilde{p}-1}}\mE|\u^\top\W_{X}|^k$ for $k\ge 1$.


\begin{thm}\label{thm.1}
Let $\underline{c}_{Z,(l)}=\lambda_{\min}(\Sigma_{Z,(l)})$ and $\underline{c}_Z=\min_l\underline{c}_{Z,(l)}$. If Assumptions 1, 2, and 4-5 hold, then for any $1\le l\le M_2$,
	\begin{align*}
		\|\wh{\btheta}_{(l)}-\btheta_{(l)}^*\|=O_p\left( \underline{c}_{Z,(l)}^{-1/2}\left(\bar{\sigma}\sqrt{\frac{q_l}{n}}+\kappa_{Z,2}^{1/2}\kappa_{X,4}^{1/2}\bar{f}r_0^2+r_0\sqrt{\frac{\tilde{p}}{n}}\right)\right).
	\end{align*}
	 Furthermore,
	\begin{align*}
		\sup_{1\le l\le M_2}\|\wh{\btheta}_{(l)}-\btheta_{(l)}^*\|= O_p\left(\underline{c}_Z^{-1/2}\left(\bar{\sigma}\sqrt{\frac{\bar{q}M_2}{n}}+\kappa_{Z,2}^{1/2}\kappa_{X,4}^{1/2}\bar{f}r_0^2+r_0\sqrt{\frac{\tilde{p}+\log(M_2)}{n}}\right)\right).
	\end{align*}
\end{thm}

Theorem~\ref{thm.1} establishes the consistency of the second-stage estimator $\wh{\btheta}_{(l)}$ for each candidate model. Importantly, this result only requires the existence of correct models in the first stage. The candidate models in the second stage are allowed to be misspecified. The error rate in Theorem~\ref{thm.1} highlights a key advantage of the Neyman-orthogonal construction. The estimation error from the first-stage VaR model averaging affects the second-stage estimator only through higher-order terms, namely $r_0^2$ and $r_0\sqrt{\tilde{p}/n}$. As a result, moderate estimation errors in the first stage do not propagate linearly into the second stage, which helps stabilize the overall procedure.

The next theorem extends this result from individual models to the model averaging estimator. Suppose that the $l_0$-th candidate model in the second stage is correctly specified, and let $\w_0$ denote the weight vector that assigns unit weight to this model and zero weight to all others. Define $\bTheta_0=\bTheta(\w_0)$, where $\bTheta(\w)=\sum_{l=1}^{M_2}w_l\bTheta_{(l)}^*$ and $\bTheta_{(l)}^*=({\bm P}_{(l)}^E)^\top\btheta_{(l)}^*$. Let $\bar{c}_Z=\max_{1\le l\le M_2}\lambda_{\max}(\Sigma_{Z,(l)})$ and $r_1=n^{-1/2}\bar{q}^{1/2}M_2^{1/2}+\kappa_{Z,2}^{1/2}\kappa_{X,4}^{1/2}r_0^2+n^{-1/2}(\tilde{p}+\log(M_2))^{1/2}r_0$.
\begin{thm}
	\label{thm.2}
	Under the same conditions in Theorem~\ref{thm.1}, we have
	\begin{align*}
		\|\wh{\bm{\Theta}}(\wh{\w}_{ES})-\bTheta_0\|=O_p\left(\bar{c}_Z\underline{c}_Z^{-3/2}r_1\right).
	\end{align*}
\end{thm}
Theorem \ref{thm.2} shows that when there exist correct models in the second stage, the model averaging estimator is consistent for $\bTheta_0$. This result guarantees that the averaging step does not deteriorate the estimation accuracy relative to the correctly specified model.

We next study the behavior of the weights $\wh{\w}_{ES}$. Let $\xi_n=\inf_{\w\in\mH_{M_2}^{\calD^E}}\text{EFPE}^*(\w)$ measure the minimal excess forecast error among models outside $\calD^E$, where $\mH_{M_2}^{\calD^E}\equiv \{\w\in \mH_{M_2}:\sum_{l\in \calD^E}w_l=0\}$, $\text{EFPE}^*(\w)=\mE\{\wt{Y}_{0}^*-\wt{\Z}_{0}^\top\bTheta^*(\w)\}^2-\mE\epsilon_{0}^2$, and $\epsilon_{0}=\wt{Y}_0^*-\wt{\Z}_{0}^\top\bTheta_0$.
\begin{thm}
	\label{thm.3}
	Under the same conditions in Theorem~\ref{thm.1}, we have
	\begin{align*}
		1-\sum_{l\in \calD^E}\wh{w}_{ES,(l)}=O_p(\xi_n^{-1/2}\bar{c}_Z^{1/2}r_1+\xi_n^{-1/2}\bar{c}_X^{1/4}r_0^{1/2}+n^{-1/4}\xi_n^{-1/2}M_2^{1/2}).
	\end{align*}
\end{thm}
Theorem~\ref{thm.3} shows that the total weight assigned to incorrect models converges to zero at a rate determined by $\xi_n$ and the complexity of the candidate model set. In particular, when only a single second-stage model is correctly specified, the corresponding weight converges to one. In this case, the proposed ESMA procedure asymptotically reduces to model selection.

Subsequently, we study the theoretical properties of the proposed method when all candidate models are misspecified, that is, $\calD^Q=\calD^E=\emptyset$. In this case, consistency toward a true finite-dimensional parameter is no longer attainable, and the relevant benchmark becomes risk optimality.

Recall that the first-stage model averaging estimator is obtained by minimizing the expected check loss. Let $R_{VaR}(\w)=\mE\rho_\tau\{Y_i-\wt{\X}_i^\top\wh{\V}(\w)\}$ and the population-optimal weight vector $\w_Q^*=(w_{Q,1}^*,\ldots,w_{Q,M_1}^*)=\argmin_{\w\in\mH_{M_1}}R_{VaR}(\w)$, where $\mH_{M_1}=\{\w\in [0,1]^{M_1}: \|\w\|_1=1\}$.

\noindent\textbf{Assumption 6} The weights in the first stage satisfy $\|\wh{\w}_Q-\w_Q^*\|=O_p(r_n^w)$.

Assumption 6 requires the estimated first-stage weights $\wh{\w}_Q$ to converge to $\w_Q^*$ at rate $r_n^w$. This is a mild condition, since the difference between the cross-validation criterion $CV_Q(\w)$ and its population counterpart $R_{VaR}(\w)$ is $o_p(1)$. A closely related result for general loss functions, including the check loss, is established in Theorem 4 of \citet{Yu2025Unified}.

Let $\V_{(m)}^*=({\bm P}_{(m)}^Q)^\top\bnu_{(m)}^*$ and $\V^*(\w_Q^*)=\sum_{m=1}^{M_1}w_{Q,m}^*\V_{(m)}^*$. By Assumptions 3 and 6, the first-stage model averaging coefficient satisfies $\|\wh{\V}(\wh{\w}_Q)-\V^*(\w_Q^*)\|=O_p(r^Q+M_1^{1/2}\tilde{p}^{1/2}r_n^w)$. To clarify the target of the second-stage regression under first-stage
misspecification, let
$
q_\tau^\dagger(\X)
=
\widetilde{\X}^{\top}\V^*(\w_Q^*)
$
denote the pseudo-true VaR induced by the first-stage candidate models, and
define
$
\ES_\tau^\dagger(\X)
=
\mathbb{E}\left[
\widetilde{Y}\{\V^*(\w_Q^*)\}
\mid \X
\right].
$
When $q_\tau^\dagger(\X)=\VaR_\tau(Y\mid\X)$ almost surely,
$\ES_\tau^\dagger(\X)$ coincides with the true conditional expected
shortfall. Under misspecification, however, this equality need not hold,
and the second-stage regression targets the pseudo-ES
$\ES_\tau^\dagger(\X)$ rather than the true ES. The discrepancy can be quantified locally. For
$
m_\tau(v,\X)
=
v+\tau^{-1}
\mathbb{E}\left[
(Y-v)\mathbb{I}(Y\le v)
\mid\X
\right],
$
let $q_\tau(\X)=\VaR_\tau(Y\mid\X)$. Then
\[
m_\tau(v,\X)-m_\tau\{q_\tau(\X),\X\}
=
\int_{q_\tau(\X)}^v
\left\{
1-\tau^{-1}F_{Y\mid\X}(u\mid\X)
\right\}\mathrm{d}u.
\]
Under the Lipschitz condition in Assumption~2,
\[
\left|
m_\tau(v,\X)-\ES_\tau(Y\mid\X)
\right|
\le
\frac{\bar f}{2\tau}
\left|v-q_\tau(\X)\right|^2.
\]
Thus, a small first-stage approximation error affects the conditional
target only at the second order. If, however,
$q_\tau^\dagger(\X)$ remains separated from the true conditional
quantile, $\ES_\tau^\dagger(\X)$ need not converge to the true conditional
ES. In this case, the first optimality result in
Theorem~\ref{thm.opt} should be interpreted with respect to the
pseudo-ES risk rather than the oracle risk associated with the true ES. Let $\wt{\varepsilon}_i=Y_i-\wt{\X}_i^\top\V^*(\w_Q^*)$.

\noindent\textbf{Assumption 7} The conditional distribution function $F_{\wt{\varepsilon}_i|\wt{\X}_i,\wt{\Z}_i}$ given $\wt{\X}_i$ and $\wt{\Z}_i$ is continuously differentiable and satisfies $|F_{\wt{\varepsilon}_i|\wt{\X}_i,\wt{\Z}_i}(t)-F_{\wt{\varepsilon}_i|\wt{\X}_i,\wt{\Z}_i}(0)|\le \bar{f}_Q|t|$ for some positive constant $\bar{f}_Q$ and all $t\in\mR$.

Assumption 7 is similar to Assumption 2. Let $r_2=n^{-1/2}\bar{q}^{1/2}M_2^{1/2}+\kappa_{Z,2}^{1/2}\kappa_{X,4}^{1/2}\bar{f}_Q(r^Q+M_1^{1/2}\tilde{p}^{1/2}r_n^w)^2+n^{-1/2}(\tilde{p}+\log(M_2))^{1/2}(r^Q+M_1^{1/2}\tilde{p}^{1/2}r_n^w)$. Under Assumptions 3 and 7, an argument similar to that in Theorem 1 yields $\max_{1\le l\le M_2}\|\wh{\btheta}_{(l)}-\btheta_{(l)}^\dagger\|=O_p(\underline{c}_Z^{-1/2}r_2)$, where $\btheta_{(l)}^\dagger=\argmin_{\btheta_{(l)}}\mE\{\wt{Y}_i(\V^*(\w_Q^*))-\Z_{(l),i}^\top\btheta_{(l)}\}^2$. Let $\epsilon_{(l),i}^\dagger=\wt{Y}_i(\V^*(\w_Q^*))-\Z_{(l),i}^\top\btheta_{(l)}^\dagger$.

\noindent\textbf{Assumption 8} The moments $\mE(\epsilon_{(l),i}^\dagger)^4$ are uniformly bounded.

Assumption 8 imposes an boundedness condition for the residuals $\epsilon_{(l),i}^\dagger$. Let $R_{ES}^\dagger(\w)=\mE[\{\wt{Y}_0(\V^*(\w_Q^*))-\wt{\Z}_0^\top\wh{\bTheta}(\w)\}^2|\calD_n]$ denote the expected shortfall risk given  $\calD_n$ and $R_{ES}^{\dagger*}(\w)=\mE[\{\wt{Y}_0(\V^*(\w_Q^*))-\wt{\Z}_0^\top\bTheta^\dagger(\w)\}^2]$, where $\bTheta^\dagger(\w)=\sum_{l=1}^{M_2}w_l\bTheta_{(l)}^\dagger$ and $\bTheta_{(l)}^\dagger=({\bm P}_{(l)}^E)^\top\btheta_{(l)}^\dagger$. Let $\xi_n^\dagger=\min_{\w \in\mH_{M_2}}R_{ES}^{\dagger*}(\w)$ denote the minimum achievable second-stage risk.

\noindent\textbf{Assumption 9} (i) $n^{-1/2}(\xi_n^\dagger)^{-1}M_2=o_p(1)$; (ii) $(\xi_n^\dagger)^{-1}\underline{c}_Z^{-1/2}\bar{c}_Z^{1/2}r_2=o_p(1)$; (iii) $(\xi_n^\dagger)^{-1}\bar{c}_X^{1/2}(r^Q+M_1^{1/2}\tilde{p}^{1/2}r_n^w)=o_p(1)$. 

Assumption 9 imposes rate conditions that balance the sample size, the number of candidate models, and the dimensions of the predictors. Let $R_{ES}^\ddagger(\w)=\mE[\{\wt{Y}_0^*-\wt{\Z}_0^\top\wh{\bTheta}(\w)\}^2|\calD_n]$ denote the ideal expected shortfall risk when the first-stage coefficient is known and $\xi_n^\ddagger=\min_{\w \in\mH_{M_2}}\mE[\{\wt{Y}_0^*-\wt{\Z}_0^\top\bTheta^\dagger(\w)\}^2]$.

\noindent\textbf{Assumption 10} (i) $n^{-1/2}(\xi_n^\ddagger)^{-1}M_2=o_p(1)$; (ii) $(\xi_n^\ddagger)^{-1}\underline{c}_Z^{-1/2}\bar{c}_Z^{1/2}r_2=o_p(1)$; (iii) $(\xi_n^\ddagger)^{-1}\bar{c}_X^{1/2}(r^Q+M_1^{1/2}\tilde{p}^{1/2}r_n^w)=o_p(1)$; (iv) $(\xi_n^\ddagger)^{-1}\mE(\wt{Y}_0(\V^*(\w_Q^*))-\wt{Y}_0^*)^2=o(1)$.

Assumptions 10(i)-(iii) are similar to Assumption 9. Assumption 10(iv) is an additional approximation condition linking the
pseudo-response induced by the misspecified first-stage candidate models
to the oracle pseudo-response constructed from the true conditional VaR.
 The condition can hold
when the first-stage candidate set is enriched with the sample size, as in
sieve or series approximations, so that
$q_\tau^\dagger(\X)$ approaches $q_\tau(\X)$. Without
Assumption 10(iv), the optimality result below applies to
$R_{ES}^\dagger$ but does not imply optimality for the true ES target.

\begin{thm}\label{thm.opt}
	If Assumptions 1-3 and 6-9 hold, then the ESMA estimator satisfies the asymptotic optimality property
	\begin{align*}
		\frac{R_{ES}^\dagger(\wh{\w}_{ES})}{\inf_{\w\in \mH_{M_2}}R_{ES}^\dagger(\w)}=1+o_p(1).
	\end{align*}
	If Assumption 10 additionally holds, then
	\begin{align*}
		\frac{R_{ES}^\ddagger(\wh{\w}_{ES})}{\inf_{\w\in \mH_{M_2}}R_{ES}^\ddagger(\w)}=1+o_p(1).
	\end{align*}
\end{thm}
Theorem~\ref{thm.opt} shows that, when all candidate models are misspecified, the proposed ESMA procedure achieves asymptotic risk optimality. That is, the cross-validation weights asymptotically attain the lowest possible expected shortfall risk among all convex combinations of the candidate models. It is worth noting that the risk $R_{ES}^\dagger(\w)$ depends on the first-stage candidate set through $\wt{Y}_0(\V^*(\w_Q^*))$. By enriching the candidate models in the first stage, the gap between $\V^*(\w_Q^*)$ and $\V_0$ may be reduced. When $\wt{Y}_0(\V^*(\w_Q^*))$ well approximates $\wt{Y}_0^*$, ESMA weights are also asymptotically optimal with respect to the oracle risk $R_{ES}^\ddagger(\w)$.

\section{Monte Carlo Simulation}\label{chap:sim}
\subsection{Design 1}
Following the setting of \citet{Jiao2025Optimal}, we consider the DGP:
\begin{align*}
	Y_i = \alpha\sum_{j=1}^{1000}j^{-1}X_{ij}+\varepsilon_i,
\end{align*}
where $X_{i1}=-1$ and $X_{ij},\ j=2,3,\ldots$ are each i.i.d. $Normal(0,1)$ and are mutually independent of each other. We consider both homoscedastic and heteroscedastic error structures. In the homoscedastic case we set $\varepsilon_i\sim Normal(0,1)$. In the heteroscedastic case, we generate $\varepsilon_i=(\sum_{j=2}^6X_{ij}^2)\eta_i$ and $\eta_i\sim Normal(0,1)$. The quantile level is $\tau=0.05$ and 0.1. The signal strength is controlled by the parameter $\alpha$, which is chosen such that the population $R^2$
ranges from 0.1 to 0.9. We consider three sample sizes $n=100,200$ and 400. The number of candidate models is set to $M=\lfloor3n^{1/3}\rfloor$, where $\lfloor\cdot\rfloor$ is the integer part. We construct nested candidate models, that is, the first model contains the first variable $X_{i1}$, the second model contains the first and second variables $X_{i1}, X_{i2}$, and so forth. Forecasting performance is evaluated using an independent out-of-sample test set of size 100. The final prediction errors are calculated by
\begin{align*}
	\text{FPE}_1=\frac{1}{100}\sum_{i=1}^{100}(\widetilde{Y}_i^*-\widehat{\ES}_i)^2
\end{align*}
and
\begin{align*}
	\text{FPE}_2=\frac{1}{100}\sum_{i=1}^{100}(Y_i-\widehat{\ES}_i)^2\mathbb{I}(Y_i\le \VaR(Y_i|\X_i)),
\end{align*}
where $\widetilde{Y}_i^*=\alpha^{-1}(Y_i-\VaR(Y_i|\X_i))\mathbb{I}(Y_i\le \VaR(Y_i|\X_i))+\VaR(Y_i|\X_i)$ and $\VaR(Y_i|\X_i)$ is the true VaR. \endnote{ While $\text{FPE}_1$ corresponds to the risk function studied in Section~\ref{chap:theory}, $\text{FPE}_2$ corresponds to an alternative risk $\mE[(Y_i-\ES_i)^2|Y_i\le \VaR(Y_i|\X_i),\X_i]$ which is minimized at the true conditional expected shortfall $\ES_i$.} The excess final prediction errors are defined as 
\begin{align*}
	\text{EFPE}_1=\frac{1}{100}\sum_{i=1}^{100}(\widetilde{Y}_i^*-\widehat{\ES}_i)^2-\frac{1}{100}\sum_{i=1}^{100}(\widetilde{Y}_i^*-\ES_i)^2
\end{align*}
and
\begin{align*}
	\text{EFPE}_2=\frac{1}{100}\sum_{i=1}^{100}(Y_i-\widehat{\ES}_i)^2\mathbb{I}(Y_i\le \VaR(Y_i|\X_i))-\frac{1}{100}\sum_{i=1}^{100}(Y_i-\ES_i)^2\mathbb{I}(Y_i\le \VaR(Y_i|\X_i)),
\end{align*}
where $\ES_i$ is the true expected shortfall, which is available in closed form and reported in Section S.1 of the Online Supplement. We set $K=10$ for ESMA. The competing methods include full model (FM) by \citet{barendseefficiently}, robust regression (Robust) by \citet{He2023}, high-dimensional penalized regression (HighDim) by \citet{Zhang2025}, and joint model averaging (JointMA) by \citet{Jiao2025Optimal}. All results are averaged over 100 Monte Carlo replications. To save space we only show the results of homoscedastic cases in Tables \ref{table.1}-\ref{table.2}. The results of  heteroscedastic cases can be found in the Section S.3 of the Online Supplement. 

Across all designs, sample sizes, and values of $R^2$, ESMA consistently achieves the smallest or near-smallest EFPE among competing methods. This advantage is particularly pronounced in moderate-to-high signal regimes ($R^2\ge 0.3$) and becomes more stable as the sample size increases. First, when evaluated under EFPE$_1$, which corresponds to the risk function studied in the theoretical analysis, ESMA uniformly outperforms benchmark methods in both homoscedastic and heteroscedastic settings. This empirical evidence supports the theoretical optimality results established in Section~\ref{chap:theory}. Second, under EFPE$_2$, which corresponds to a different conditional mean squared error criterion, ESMA continues to perform favorably. Although EFPE$_2$ is not the explicit optimization target of ESMA, its performance remains competitive and often superior to existing methods, indicating that the proposed approach does not overfit a specific loss function, but instead yields stable predictions across different evaluation metrics. Third, the behavior of HighDim and Robust methods differs markedly across signal regimes. HighDim performs competitively only in very weak-signal settings, where sparsity assumptions approximately hold, but its performance degrades as $R^2$ increases, indicating sensitivity of model selection. In contrast, the Robust method benefits from stronger signal strength and generally improves as $R^2$ increases, reflecting the fact that robust loss functions become more effective once the underlying conditional structure is reasonably well identified. Nevertheless, despite this improvement, Robust remains consistently dominated by ESMA in most scenarios. This suggests that while robustness is helpful for stabilizing estimation in high-signal regimes, it does not fully address model uncertainty, which is crucial for achieving optimal ES prediction. Fourth, JointMA is designed to optimize a joint criterion, it often exhibits substantially larger EFPE values, especially for large $R^2$ and larger $n$ when $\tau=0.05$. In these cases, JointMA becomes numerically unstable, leading to extremely large prediction errors, whereas ESMA remains well behaved. 

\begin{table}[!ht]
	\centering
	
	\caption{EFPE$_1$ under the homoscedastic setting in Design 1}\label{table.1}
	\renewcommand{\arraystretch}{0.6}
	\small 
	\setlength{\tabcolsep}{6.0pt} 
	\begin{threeparttable}
		\begin{tabular}{lllllllllll}
			\toprule
			$n$ & Method & 0.1 & 0.2 & 0.3 & 0.4 & 0.5 & 0.6 & 0.7 & 0.8 & 0.9 \\ 
			\hline
			\multicolumn{11}{c}{$\tau = 0.05$} \\ 
			\hline
			100 & ESMA & 0.306 & 0.435 & \textbf{0.500} & \textbf{0.519} & \textbf{0.551} & \textbf{0.747} & \textbf{1.006} & \textbf{1.283} & 2.363 \\ 
			& FM & 1.041 & 1.149 & 1.122 & 1.102 & 1.124 & 1.360 & 1.375 & 1.492 & 2.275 \\ 
			& Robust & 1.007 & 1.101 & 1.067 & 1.034 & 1.049 & 1.269 & 1.281 & 1.448 & \textbf{2.135} \\ 
			& HighDim & \textbf{0.240} & \textbf{0.391} & 0.601 & 0.737 & 0.950 & 1.265 & 1.505 & 1.817 & 3.025 \\ 
			& JointMA & 0.388 & 0.589 & 1.060 & 1.162 & 1.824 & 2.665 & 42.467 & 7.480 & 17.771 \\ 
			200 & ESMA & \textbf{0.181} & \textbf{0.264} & \textbf{0.295} & \textbf{0.371} & \textbf{0.427} & \textbf{0.431} & \textbf{0.587} & \textbf{0.852} & 1.727 \\ 
			& FM & 0.739 & 0.808 & 0.674 & 0.734 & 0.750 & 0.776 & 0.917 & 1.040 & 1.616 \\ 
			& Robust & 0.697 & 0.777 & 0.644 & 0.709 & 0.729 & 0.721 & 0.877 & 0.871 & \textbf{1.329} \\ 
			& HighDim & 0.201 & 0.330 & 0.439 & 0.559 & 0.613 & 0.631 & 0.841 & 1.085 & 1.842 \\ 
			& JointMA & 0.247 & 0.442 & 0.581 & 0.580 & 1.134 & 1.753 & 2.578 & 10.202 & 20.415 \\ 
			400 & ESMA & \textbf{0.115} & \textbf{0.158} & \textbf{0.218} & \textbf{0.213} & \textbf{0.298} & \textbf{0.329} & \textbf{0.431} & 0.692 & 1.290 \\ 
			& FM & 0.462 & 0.384 & 0.469 & 0.409 & 0.503 & 0.455 & 0.523 & 0.807 & 1.215 \\ 
			& Robust & 0.409 & 0.318 & 0.405 & 0.317 & 0.398 & 0.382 & 0.436 & \textbf{0.646} & \textbf{1.035} \\ 
			& HighDim & 0.130 & 0.229 & 0.327 & 0.347 & 0.400 & 0.403 & 0.505 & 0.804 & 1.232 \\ 
			& JointMA & 0.120 & 0.233 & 0.304 & 0.614 & 0.643 & 1.264 & 2.022 & 5.505 & 23.578 \\ 
			\bottomrule
			\multicolumn{11}{c}{$\tau = 0.1$} \\ 
			\hline
			100 & ESMA & 0.205 & \textbf{0.273} & \textbf{0.296} & \textbf{0.407} & \textbf{0.492} & \textbf{0.607} & \textbf{0.774} & \textbf{1.214} & 2.268 \\ 
			& FM & 0.780 & 0.829 & 0.735 & 0.877 & 0.913 & 0.915 & 1.062 & 1.401 & 2.258 \\ 
			& Robust & 0.770 & 0.785 & 0.731 & 0.796 & 0.916 & 0.880 & 1.008 & 1.226 & \textbf{1.884} \\ 
			& HighDim & \textbf{0.190} & 0.327 & 0.431 & 0.581 & 0.693 & 0.784 & 0.969 & 1.388 & 2.384 \\ 
			& JointMA & 0.265 & 0.417 & 0.490 & 0.820 & 1.154 & 1.976 & 4.015 & 9.136 & 19.104 \\ 
			200 & ESMA & \textbf{0.145} & \textbf{0.164} & \textbf{0.242} & \textbf{0.232} & \textbf{0.312} & \textbf{0.426} & \textbf{0.554} & 0.753 & 1.537 \\ 
			& FM & 0.436 & 0.461 & 0.510 & 0.468 & 0.518 & 0.575 & 0.684 & 0.782 & 1.427 \\ 
			& Robust & 0.394 & 0.371 & 0.392 & 0.407 & 0.456 & 0.509 & 0.577 & \textbf{0.690} & \textbf{1.260} \\ 
			& HighDim & 0.181 & 0.222 & 0.337 & 0.338 & 0.409 & 0.531 & 0.650 & 0.791 & 1.558 \\ 
			& JointMA & 0.202 & 0.235 & 0.398 & 0.508 & 0.796 & 1.376 & 2.795 & 5.292 & 14.296 \\ 
			400 & ESMA & \textbf{0.073} & \textbf{0.110} & \textbf{0.170} & \textbf{0.171} & \textbf{0.251} & \textbf{0.270} & 0.385 & 0.496 & 1.139 \\ 
			& FM & 0.269 & 0.258 & 0.295 & 0.293 & 0.334 & 0.335 & 0.415 & 0.534 & 1.085 \\ 
			& Robust & 0.195 & 0.196 & 0.238 & 0.215 & 0.281 & 0.276 & \textbf{0.329} & \textbf{0.462} & \textbf{0.972} \\ 
			& HighDim & 0.106 & 0.156 & 0.216 & 0.220 & 0.304 & 0.314 & 0.426 & 0.521 & 1.107 \\ 
			& JointMA & 0.088 & 0.167 & 0.243 & 0.360 & 0.603 & 1.101 & 1.900 & 3.628 & 10.767 \\ 
			\bottomrule
		\end{tabular}
		\begin{tablenotes}\footnotesize
			\item[] Notes: Columns 3--11 report results for $R^2=0.1,0.2,\ldots,0.9$.  Boldface indicates the best-performing method.
		\end{tablenotes}
	\end{threeparttable}
\end{table}

\begin{table}[!ht]
	\centering
	\caption{EFPE$_2$ under the homoscedastic setting in Design 1}\label{table.2}
	\renewcommand{\arraystretch}{0.6}
	\small 
	\setlength{\tabcolsep}{6.0pt} 
	\begin{threeparttable}
		\begin{tabular}{lllllllllll}
			\toprule
			$n$ & Method & $R^2=0.1$ & 0.2 & 0.3 & 0.4 & 0.5 & 0.6 & 0.7 & 0.8 & 0.9 \\ 
			\hline
			\multicolumn{11}{c}{$\tau = 0.05$} \\ 
			\hline
			100 & ESMA & 0.016 & 0.021 & \textbf{0.025} & \textbf{0.025} & \textbf{0.026} & \textbf{0.035} & \textbf{0.046} & \textbf{0.058} & 0.138 \\ 
			& FM & 0.052 & 0.054 & 0.055 & 0.054 & 0.050 & 0.069 & 0.070 & 0.078 & 0.127 \\ 
			& Robust & 0.049 & 0.053 & 0.053 & 0.051 & 0.048 & 0.066 & 0.065 & 0.075 & \textbf{0.118} \\ 
			& HighDim & \textbf{0.013} & \textbf{0.020} & 0.029 & 0.035 & 0.044 & 0.059 & 0.073 & 0.093 & 0.168 \\ 
			& JointMA & 0.021 & 0.036 & 0.052 & 0.052 & 0.101 & 0.131 & 1.509 & 0.373 & 0.947 \\ 
			200 & ESMA & 0.011 & \textbf{0.012} & \textbf{0.015} & \textbf{0.018} & \textbf{0.020} & \textbf{0.023} & \textbf{0.036} & 0.048 & 0.092 \\ 
			& FM & 0.041 & 0.038 & 0.031 & 0.037 & 0.033 & 0.040 & 0.053 & 0.050 & 0.081 \\ 
			& Robust & 0.038 & 0.036 & 0.031 & 0.035 & 0.034 & 0.037 & 0.048 & \textbf{0.044} & \textbf{0.065} \\ 
			& HighDim & \textbf{0.010} & 0.015 & 0.020 & 0.027 & 0.029 & 0.033 & 0.049 & 0.058 & 0.097 \\ 
			& JointMA & 0.015 & 0.017 & 0.029 & 0.027 & 0.057 & 0.082 & 0.140 & 0.415 & 1.117 \\ 
			400 & ESMA & \textbf{0.006} & \textbf{0.008} & \textbf{0.010} & \textbf{0.012} & \textbf{0.016} & \textbf{0.019} & \textbf{0.021} & \textbf{0.029} & 0.070 \\ 
			& FM & 0.023 & 0.020 & 0.025 & 0.023 & 0.028 & 0.024 & 0.027 & 0.037 & 0.069 \\ 
			& Robust & 0.021 & 0.016 & 0.023 & 0.018 & 0.022 & \textbf{0.019} & \textbf{0.021} & 0.030 & \textbf{0.057} \\ 
			& HighDim & 0.007 & 0.011 & 0.016 & 0.017 & 0.022 & 0.021 & 0.025 & 0.035 & 0.067 \\ 
			& JointMA & 0.007 & 0.012 & 0.014 & 0.027 & 0.032 & 0.059 & 0.088 & 0.227 & 1.543 \\ 
			\bottomrule
			\multicolumn{11}{c}{$\tau = 0.1$} \\ 
			\hline
			100 & ESMA & 0.022 & \textbf{0.029} & \textbf{0.034} & \textbf{0.036} & \textbf{0.047} & \textbf{0.059} & \textbf{0.083} & \textbf{0.118} & 0.242 \\ 
			& FM & 0.078 & 0.086 & 0.078 & 0.077 & 0.093 & 0.083 & 0.117 & 0.130 & 0.247 \\ 
			& Robust & 0.078 & 0.078 & 0.074 & 0.073 & 0.094 & 0.085 & 0.105 & 0.120 & \textbf{0.201} \\ 
			& HighDim & \textbf{0.021} & 0.031 & 0.044 & 0.055 & 0.067 & 0.073 & 0.098 & 0.133 & 0.257 \\ 
			& JointMA & 0.028 & 0.043 & 0.056 & 0.081 & 0.121 & 0.184 & 0.403 & 0.885 & 2.241 \\ 
			200 & ESMA & \textbf{0.016} & \textbf{0.016} & \textbf{0.027} & \textbf{0.024} & \textbf{0.031} & \textbf{0.038} & \textbf{0.055} & 0.078 & 0.158 \\ 
			& FM & 0.044 & 0.043 & 0.052 & 0.049 & 0.052 & 0.055 & 0.070 & 0.082 & 0.151 \\ 
			& Robust & 0.039 & 0.035 & 0.041 & 0.041 & 0.046 & 0.048 & 0.059 & \textbf{0.074} & \textbf{0.133} \\ 
			& HighDim & 0.019 & 0.023 & 0.036 & 0.031 & 0.040 & 0.046 & 0.067 & 0.081 & 0.163 \\ 
			& JointMA & 0.021 & 0.024 & 0.042 & 0.050 & 0.084 & 0.143 & 0.299 & 0.544 & 1.261 \\ 
			400 & ESMA & \textbf{0.008} & \textbf{0.012} & \textbf{0.014} & \textbf{0.018} & \textbf{0.023} & \textbf{0.027} & \textbf{0.033} & 0.051 & 0.114 \\ 
			& FM & 0.026 & 0.027 & 0.027 & 0.032 & 0.035 & 0.035 & 0.039 & 0.052 & 0.108 \\ 
			& Robust & 0.019 & 0.020 & 0.023 & 0.024 & 0.028 & 0.028 & \textbf{0.033} & \textbf{0.045} & \textbf{0.098} \\ 
			& HighDim & 0.011 & 0.016 & 0.019 & 0.024 & 0.029 & 0.033 & 0.037 & 0.052 & 0.109 \\ 
			& JointMA & 0.009 & 0.016 & 0.022 & 0.034 & 0.054 & 0.102 & 0.158 & 0.352 & 0.932 \\ 
			\bottomrule
		\end{tabular}
		\begin{tablenotes}\footnotesize
			\item[] Notes: Columns 3--11 report results for $R^2=0.1,0.2,\ldots,0.9$.  Boldface indicates the best-performing method.
		\end{tablenotes}
	\end{threeparttable}
\end{table}

Finally, to highlight the computational advantage of the proposed ESMA method over JointMA, we report the average computing time under the homoscedastic setting in Table~\ref{table.3}. Across all configurations, ESMA requires approximately one tenth of the computing time of JointMA. Moreover, as the sample size $n$ increases, the computational cost of JointMA grows rapidly, reflecting the substantial overhead induced by the jackknife procedure used in its weight selection criterion.

\begin{table}[!ht]
	\centering
	\caption{Average computing time in seconds under homoscedastic setting in Design 1}
	\label{table.3}
	\renewcommand{\arraystretch}{0.6}
	\small 
	\setlength{\tabcolsep}{5pt}
		\begin{threeparttable}
			\begin{tabular}{llllllllllll}
				\toprule
				$\tau$ & $n$ & Method & 0.1 & 0.2 & 0.3 & 0.4 & 0.5 & 0.6 & 0.7 & 0.8 & 0.9 \\ \midrule
				0.05 & 100 & ESMA & 0.63 & 0.35 & 0.33 & 0.33 & 0.33 & 0.33 & 0.34 & 0.34 & 0.35 \\ 
				~ & ~ & JointMA & 38.54 & 38.40 & 38.70 & 38.81 & 38.93 & 39.09 & 39.35 & 39.52 & 39.85 \\ 
				~ & 200 & ESMA & 0.47 & 0.48 & 0.48 & 0.48 & 0.49 & 0.49 & 0.49 & 0.50 & 0.51 \\ 
				~ & ~ & JointMA & 145.83 & 145.94 & 146.06 & 146.67 & 147.31 & 148.16 & 148.55 & 149.53 & 149.92 \\ 
				~ & 400 & ESMA & 0.70 & 0.71 & 0.72 & 0.72 & 0.73 & 0.74 & 0.74 & 0.75 & 0.77 \\ 
				~ & ~ & JointMA & 517.70 & 518.49 & 519.00 & 520.71 & 521.16 & 525.57 & 527.41 & 526.69 & 525.35 \\ 
				0.1 & 100 & ESMA & 0.77 & 0.38 & 0.35 & 0.35 & 0.36 & 0.36 & 0.36 & 0.36 & 0.36 \\ 
				~ & ~ & JointMA & 40.96 & 41.13 & 41.32 & 41.43 & 41.57 & 41.78 & 42.02 & 42.24 & 42.35 \\ 
				~ & 200 & ESMA & 0.50 & 0.51 & 0.51 & 0.52 & 0.52 & 0.53 & 0.53 & 0.53 & 0.54 \\ 
				~ & ~ & JointMA & 156.94 & 156.75 & 157.21 & 157.63 & 158.00 & 158.24 & 159.19 & 159.64 & 161.17 \\ 
				~ & 400 & ESMA & 0.78 & 0.78 & 0.78 & 0.84 & 0.82 & 0.81 & 0.82 & 0.83 & 0.83 \\ 
				~ & ~ & JointMA & 567.25 & 562.40 & 565.30 & 583.65 & 563.98 & 565.68 & 572.52 & 562.24 & 555.88 \\ \bottomrule
			\end{tabular}
			\begin{tablenotes}\footnotesize
				\item[] Note: Columns 4--12 report results for $R^2=0.1,0.2,\ldots,0.9$.
			\end{tablenotes}
		\end{threeparttable}
\end{table}

\subsection{Design 2}
Following \citet{Lu2015}, we consider the DGP
\begin{align*}
	Y_i=\alpha\left\{X_{i1}+\sum_{j=2}^{25}j^{-1}\Phi(X_{ij})\right\}+\varepsilon_i,
\end{align*}
where $X_{i1}=-1$ and $\{X_{ij}\}_{j>1}$ are i.i.d. $Normal(0,1)$, mutually independent of each other. Here, $\Phi(\cdot)$ is the standard normal cumulative distribution function. In the homoscedastic setting, we generate the error term $\varepsilon_i\sim Normal(0,1)$. In the heteroscedastic situation, we set $\varepsilon_i=(0.01+\sum_{j=2}^{11}X_{ij}^2)\eta_i$ and $\eta_i\sim Normal(0,1)$. We fix the number of candidate models at $M=20$ in all cases. The $m$-th model uses $\{X_{ij}\}_{j=1}^m$ as predictors. The competing methods and evaluation procedures are identical to those in Design 1. The simulation results are reported in Tables \ref{table.4}-\ref{table.5}. In addition, we compare the computational efficiency of ESMA and JointMA in Table~\ref{tab.simu2.time}.

The empirical findings in Design 2 largely corroborate those observed in Design 1. Across both homoscedastic and heteroscedastic settings, ESMA continues to deliver the smallest or near-smallest EFPE under both EFPE$_1$ and EFPE$_2$, remaining consistently competitive across different sample sizes and values of $R^2$. This confirms that the superior performance of ESMA is not specific to linear data generating processes, but persists in more complex nonlinear environments. 

\begin{table}[!ht]
	\centering
	
		\caption{EFPE$_1$ under the homoscedastic setting in Design 2}\label{table.4}
		\renewcommand{\arraystretch}{0.6}
		\small 
		\setlength{\tabcolsep}{6.0pt}
		\begin{threeparttable}
		\begin{tabular}{lllllllllll}
			\toprule
			$n$ & Method & 0.1 & 0.2 & 0.3 & 0.4 & 0.5 & 0.6 & 0.7 & 0.8 & 0.9 \\ 
			\midrule
			\multicolumn{11}{c}{$\tau = 0.05$} \\ 
			\midrule
			100 & ESMA & 0.407 & 0.385 & 0.506 & \textbf{0.470} & \textbf{0.638} & \textbf{0.711} & \textbf{0.785} & \textbf{1.056} & \textbf{1.656} \\ 
			& FM & 1.527 & 1.511 & 1.666 & 1.569 & 1.546 & 1.537 & 1.668 & 1.729 & 2.151 \\ 
			& Robust & 1.513 & 1.488 & 1.662 & 1.542 & 1.526 & 1.535 & 1.630 & 1.725 & 2.094 \\ 
			& HighDim & \textbf{0.246} & 0.347 & 0.537 & 0.769 & 1.001 & 1.290 & 1.605 & 2.109 & 2.695 \\ 
			& JointMA & 0.292 & \textbf{0.245} & \textbf{0.403} & 0.645 & 1.205 & 2.025 & 3.932 & 8.574 & 24.783 \\ 
			200 & ESMA & 0.232 & 0.230 & \textbf{0.311} & \textbf{0.394} & \textbf{0.423} & \textbf{0.499} & \textbf{0.584} & \textbf{0.752} & 1.543 \\ 
			& FM & 0.845 & 0.830 & 0.856 & 0.867 & 0.779 & 0.891 & 1.016 & 0.969 & 1.584 \\ 
			& Robust & 0.827 & 0.802 & 0.830 & 0.846 & 0.710 & 0.859 & 0.942 & 0.877 & \textbf{1.382} \\ 
			& HighDim & 0.219 & 0.278 & 0.418 & 0.627 & 0.612 & 0.744 & 0.965 & 0.971 & 1.639 \\ 
			& JointMA & \textbf{0.201} & \textbf{0.223} & 0.323 & 0.656 & 1.090 & 2.041 & 3.960 & 8.422 & 25.355 \\ 
			400 & ESMA & 0.135 & \textbf{0.161} & \textbf{0.171} & \textbf{0.209} & \textbf{0.270} & \textbf{0.344} & 0.439 & 0.636 & 1.187 \\ 
			& FM & 0.395 & 0.362 & 0.383 & 0.411 & 0.430 & 0.494 & 0.505 & 0.633 & 1.157 \\ 
			& Robust & 0.317 & 0.309 & 0.308 & 0.309 & 0.371 & 0.400 & \textbf{0.399} & \textbf{0.474} & \textbf{0.959} \\ 
			& HighDim & \textbf{0.132} & 0.236 & 0.247 & 0.302 & 0.361 & 0.426 & 0.514 & 0.653 & 1.164 \\ 
			& JointMA & 0.142 & 0.197 & 0.285 & 0.533 & 1.005 & 1.913 & 3.900 & 8.914 & 24.984 \\ 
			\bottomrule
			\multicolumn{11}{c}{$\tau = 0.1$} \\ 
			\midrule
			100 & ESMA & 0.270 & 0.307 & 0.398 & \textbf{0.380} & \textbf{0.567} & \textbf{0.634} & \textbf{0.709} & \textbf{0.961} & \textbf{1.557} \\ 
			& FM & 1.154 & 1.120 & 1.261 & 1.268 & 1.224 & 1.315 & 1.356 & 1.426 & 1.909 \\ 
			& Robust & 1.128 & 1.089 & 1.208 & 1.133 & 1.154 & 1.221 & 1.279 & 1.374 & 1.695 \\ 
			& HighDim & 0.239 & 0.334 & 0.506 & 0.627 & 0.711 & 0.885 & 1.030 & 1.295 & 1.894 \\ 
			& JointMA & \textbf{0.204} & \textbf{0.218} & \textbf{0.385} & 0.670 & 1.308 & 2.289 & 4.440 & 9.366 & 26.107 \\ 
			200 & ESMA & \textbf{0.139} & \textbf{0.175} & \textbf{0.231} & \textbf{0.301} & \textbf{0.347} & \textbf{0.431} & \textbf{0.494} & 0.628 & 1.347 \\ 
			& FM & 0.528 & 0.561 & 0.552 & 0.594 & 0.523 & 0.644 & 0.702 & 0.726 & 1.323 \\ 
			& Robust & 0.462 & 0.494 & 0.459 & 0.499 & 0.423 & 0.561 & 0.613 & \textbf{0.597} & \textbf{1.122} \\ 
			& HighDim & 0.170 & 0.251 & 0.321 & 0.451 & 0.460 & 0.547 & 0.624 & 0.729 & 1.357 \\ 
			& JointMA & 0.148 & 0.191 & 0.300 & 0.674 & 1.205 & 2.273 & 4.407 & 9.120 & 26.342 \\ 
			400 & ESMA & \textbf{0.092} & \textbf{0.128} & \textbf{0.135} & \textbf{0.170} & \textbf{0.197} & 0.286 & 0.339 & 0.541 & 1.019 \\ 
			& FM & 0.227 & 0.232 & 0.234 & 0.243 & 0.270 & 0.340 & 0.378 & 0.513 & 0.968 \\ 
			& Robust & 0.189 & 0.190 & 0.188 & 0.185 & 0.229 & \textbf{0.276} & \textbf{0.292} & \textbf{0.443} & \textbf{0.883} \\ 
			& HighDim & 0.110 & 0.178 & 0.171 & 0.222 & 0.232 & 0.313 & 0.366 & 0.523 & 0.973 \\ 
			& JointMA & 0.107 & 0.185 & 0.315 & 0.583 & 1.122 & 2.262 & 4.402 & 9.588 & 25.852 \\ 
			\bottomrule
		\end{tabular}
		\begin{tablenotes}\footnotesize
			\item[] Notes: Columns 3--11 report results for $R^2=0.1,0.2,\ldots,0.9$.  Boldface indicates the best-performing method.
		\end{tablenotes}
	\end{threeparttable}
\end{table}

\begin{table}[h]
	\centering
	
		\caption{EFPE$_2$ under the homoscedastic setting in Design 2}
		\label{table.5}
		\renewcommand{\arraystretch}{0.6}
		\small 
		\setlength{\tabcolsep}{6.0pt}
		\begin{threeparttable}
		\begin{tabular}{lllllllllll}
			\toprule
			$n$ & Method & 0.1 & 0.2 & 0.3 & 0.4 & 0.5 & 0.6 & 0.7 & 0.8 & 0.9 \\ 
			\midrule
			\multicolumn{11}{c}{$\tau = 0.05$} \\ 
			\midrule
			100 & ESMA & 0.022 & 0.018 & 0.026 & \textbf{0.023} & \textbf{0.033} & \textbf{0.037} & \textbf{0.037} & \textbf{0.049} & \textbf{0.095} \\ 
			& FM & 0.076 & 0.078 & 0.077 & 0.081 & 0.076 & 0.081 & 0.085 & 0.084 & 0.106 \\ 
			& Robust & 0.076 & 0.076 & 0.078 & 0.078 & 0.076 & 0.081 & 0.085 & 0.084 & 0.104 \\ 
			& HighDim & \textbf{0.012} & 0.016 & 0.024 & 0.034 & 0.055 & 0.066 & 0.081 & 0.103 & 0.148 \\ 
			& JointMA & 0.015 & \textbf{0.012} & \textbf{0.018} & 0.030 & 0.060 & 0.103 & 0.197 & 0.398 & 1.322 \\ 
			200 & ESMA & \textbf{0.011} & \textbf{0.011} & \textbf{0.016} & \textbf{0.020} & \textbf{0.018} & \textbf{0.024} & \textbf{0.029} & \textbf{0.039} & 0.071 \\ 
			& FM & 0.040 & 0.038 & 0.039 & 0.049 & 0.034 & 0.042 & 0.049 & 0.049 & 0.079 \\ 
			& Robust & 0.039 & 0.039 & 0.039 & 0.045 & 0.030 & 0.043 & 0.047 & 0.046 & \textbf{0.070} \\ 
			& HighDim & \textbf{0.011} & 0.014 & 0.022 & 0.031 & 0.030 & 0.034 & 0.043 & 0.052 & 0.082 \\ 
			& JointMA & \textbf{0.011} & \textbf{0.011} & 0.017 & 0.032 & 0.055 & 0.100 & 0.214 & 0.431 & 1.210 \\ 
			400 & ESMA & \textbf{0.006} & \textbf{0.007} & \textbf{0.009} & \textbf{0.010} & \textbf{0.014} & \textbf{0.016} & 0.024 & 0.027 & 0.052 \\ 
			& FM & 0.019 & 0.018 & 0.019 & 0.019 & 0.022 & 0.022 & 0.029 & 0.029 & 0.053 \\ 
			& Robust & 0.015 & 0.015 & 0.015 & 0.015 & 0.018 & 0.019 & \textbf{0.023} & \textbf{0.022} & \textbf{0.045} \\ 
			& HighDim & \textbf{0.006} & 0.010 & 0.012 & 0.014 & 0.018 & 0.021 & 0.028 & 0.030 & 0.053 \\ 
			& JointMA & 0.006 & 0.008 & 0.014 & 0.022 & 0.048 & 0.093 & 0.179 & 0.390 & 1.235 \\ 
			\bottomrule
			\multicolumn{11}{c}{$\tau = 0.1$} \\ 
			\midrule
			100 & ESMA & 0.027 & 0.028 & 0.038 & \textbf{0.039} & \textbf{0.053} & \textbf{0.060} & \textbf{0.062} & \textbf{0.098} & \textbf{0.189} \\ 
			& FM & 0.116 & 0.116 & 0.123 & 0.128 & 0.124 & 0.126 & 0.126 & 0.143 & 0.212 \\ 
			& Robust & 0.115 & 0.110 & 0.121 & 0.112 & 0.121 & 0.122 & 0.121 & 0.139 & \textbf{0.189} \\ 
			& HighDim & 0.025 & 0.032 & 0.049 & 0.068 & 0.073 & 0.092 & 0.092 & 0.133 & 0.199 \\ 
			& JointMA & \textbf{0.021} & \textbf{0.022} & \textbf{0.036} & 0.070 & 0.120 & 0.226 & 0.411 & 0.934 & 2.765 \\ 
			200 & ESMA & \textbf{0.014} & \textbf{0.016} & \textbf{0.023} & \textbf{0.033} & \textbf{0.031} & \textbf{0.042} & \textbf{0.047} & 0.065 & 0.122 \\ 
			& FM & 0.054 & 0.055 & 0.052 & 0.063 & 0.050 & 0.064 & 0.066 & 0.077 & 0.121 \\ 
			& Robust & 0.046 & 0.048 & 0.045 & 0.049 & 0.042 & 0.054 & 0.055 & \textbf{0.063} & \textbf{0.100} \\ 
			& HighDim & 0.017 & 0.023 & 0.031 & 0.047 & 0.044 & 0.055 & 0.061 & 0.074 & 0.124 \\ 
			& JointMA & 0.016 & 0.019 & 0.034 & 0.066 & 0.114 & 0.239 & 0.476 & 0.951 & 2.445 \\ 
			400 & ESMA & \textbf{0.008} & \textbf{0.011} & \textbf{0.015} & \textbf{0.015} & \textbf{0.022} & 0.028 & 0.039 & 0.045 & 0.090 \\ 
			& FM & 0.020 & 0.021 & 0.025 & 0.024 & 0.030 & 0.032 & 0.040 & 0.044 & 0.088 \\ 
			& Robust & 0.017 & 0.019 & 0.020 & 0.017 & 0.025 & \textbf{0.026} & \textbf{0.032} & \textbf{0.039} & \textbf{0.085} \\ 
			& HighDim & 0.009 & 0.016 & 0.018 & 0.020 & 0.025 & 0.031 & 0.040 & 0.044 & 0.088 \\ 
			& JointMA & 0.010 & 0.016 & 0.031 & 0.054 & 0.116 & 0.213 & 0.442 & 0.912 & 2.505 \\ 
			\bottomrule
		\end{tabular}
		\begin{tablenotes}\footnotesize
			\item[] Notes: Columns 3--11 report results for $R^2=0.1,0.2,\ldots,0.9$.  Boldface indicates the best-performing method.
		\end{tablenotes}
	\end{threeparttable}
\end{table}

\begin{table}[!ht]
	\centering

		\caption{Average computing time in seconds under homoscedastic setting in Design 2}
		\label{tab.simu2.time}
		\renewcommand{\arraystretch}{0.6}
		\small 
		\setlength{\tabcolsep}{5pt}
	\begin{threeparttable}
		\begin{tabular}{llllllllllll}
			\toprule
			$\tau$ & $n$ & Method & 0.1 & 0.2 & 0.3 & 0.4 & 0.5 & 0.6 & 0.7 & 0.8 & 0.9 \\ \midrule
			0.05 & 100 & ESMA & 1.25 & 0.66 & 0.64 & 0.64 & 0.66 & 0.65 & 0.65 & 0.67 & 0.67 \\ 
			~ & ~ & JointMA & 65.37 & 62.86 & 61.87 & 61.49 & 61.15 & 60.44 & 60.40 & 60.35 & 60.27 \\ 
			~ & 200 & ESMA & 0.66 & 0.66 & 0.67 & 0.67 & 0.67 & 0.68 & 0.69 & 0.70 & 0.70 \\ 
			~ & ~ & JointMA & 186.63 & 179.05 & 175.20 & 171.55 & 170.14 & 169.67 & 170.08 & 169.99 & 170.10 \\ 
			~ & 400 & ESMA & 0.73 & 0.76 & 0.74 & 0.75 & 0.75 & 0.73 & 0.71 & 0.70 & 0.71 \\ 
			~ & ~ & JointMA & 576.16 & 558.62 & 538.34 & 522.76 & 504.43 & 488.49 & 474.12 & 461.83 & 482.59 \\ 
			0.1 & 100 & ESMA & 1.00 & 0.58 & 0.56 & 0.56 & 0.57 & 0.57 & 0.57 & 0.58 & 0.59 \\ 
			~ & ~ & JointMA & 55.74 & 54.23 & 53.55 & 53.25 & 53.53 & 53.34 & 53.31 & 53.46 & 53.43 \\ 
			~ & 200 & ESMA & 0.58 & 0.58 & 0.59 & 0.59 & 0.60 & 0.61 & 0.60 & 0.62 & 0.61 \\ 
			~ & ~ & JointMA & 159.16 & 153.82 & 151.56 & 150.33 & 150.85 & 150.21 & 150.23 & 151.65 & 151.25 \\ 
			~ & 400 & ESMA & 0.65 & 0.66 & 0.66 & 0.67 & 0.67 & 0.69 & 0.69 & 0.71 & 0.70 \\ 
			~ & ~ & JointMA & 499.17 & 477.34 & 462.68 & 457.52 & 454.89 & 456.35 & 458.49 & 455.90 & 453.34 \\ \bottomrule
		\end{tabular}
		\begin{tablenotes}\footnotesize
			\item[] Note: Columns 4--12 report results for $R^2=0.1,0.2,\ldots,0.9$.
		\end{tablenotes}
	\end{threeparttable}
\end{table}

\section{Real Data Analysis}\label{chap:empirical}
\subsection{U.S. Stock Return}
Measuring and forecasting downside risk in equity markets is a fundamental task for investors, risk managers, and regulators. In this context, expected shortfall has emerged as a key risk metric for assessing extreme losses beyond value-at-risk. We apply the proposed method to forecast the expected shortfall of stock returns. The data are obtained from \citet{Welch2007Comprehensive} and updated through December 2020.\endnote{Data source: \url{https://www.ivo-welch.org/professional/goyal-welch/}} Similar data sets have been widely used in the model averaging literature; see, for example, \citet{Lu2015} and \citet{Sun2021}.

We use monthly observations from January 1950 to December 2020, yielding a total of $T=852$ observations. The dependent variable $y$ is the excess stock return, defined as the difference between the continuously compounded return on the S\&P 500 index (including dividends) and the risk-free rate. The set of predictors includes stock variance, long-term return, dividend yield, dividend-price ratio, Treasury bill rate, smoothed earnings-price ratio, inflation, earnings-price ratio, long-term yield, the lagged dependent variable, book-to-market ratio, default return spread, default yield spread, and net equity expansion. The predictors are ranked according to the absolute value of their correlation with the dependent variable, resulting in 15 nested candidate models with a fixed intercept term.

Using fixed in-sample sizes of $T_1 = 100, 200,$ and $400$, we construct one-step-ahead forecasts of the conditional VaR and ES of stock returns.  Since the true expected shortfall and value-at-risk are unobservable in real data, we assess forecast performance using the joint loss function used in \citet{Patton2019Dynamic}:
\begin{align*}
	L_\tau(y,v,e)=-\frac{1}{\tau e}(v-y)\mathbb{I}(y\le v)+\frac{v}{e}+\log(e)-1,
\end{align*}
where $v$ and $e$ are the VaR and ES, respectively. The prediction error is defined as
\begin{align*}
	\frac{1}{T-T_1}\sum_{t=T_1+1}^T\{L_\tau(y_t,v_t,e_t)-L_\tau^{\mbox{min}}(y_t,v_t,e_t)\},
\end{align*}
where $v_t$ and $e_t$ are the VaR and ES at time $t$, and $L_\tau^{\mbox{min}}(y_t,v_t,e_t)$ denotes the minimum error among all competing methods. We implement both rolling and recursive window schemes to evaluate the stability of the proposed method under different information updating mechanisms. The rolling window uses a fixed-length sample, whereas the recursive window employs an expanding estimation window. 

Table~\ref{tab.case1} reports the out-of-sample prediction errors under rolling and recursive estimation windows for different quantile levels and in-sample sizes. Several clear patterns emerge. First, ESMA shows the most stable performance across both window schemes. Under the rolling window, ESMA performs competitively and improves steadily as the in-sample size increases. Under the recursive window, ESMA consistently achieves the smallest prediction errors across all cases, indicating that it can effectively use accumulated information without overfitting. Second, JointMA occasionally performs well under the rolling window, especially when the in-sample size $T_1$ is small. One possible explanation is that the evaluation criterion $L_\tau(y,v,e)$ coincides with the loss function optimized by JointMA. This alignment between the evaluation measure and the estimation objective can be beneficial in small samples. However, this advantage weakens as the in-sample size increases and under the recursive window, where the expanding sample makes joint loss optimization more complex and reduces the variance advantage relative to ESMA. Third, HighDim performs worse than both ESMA and JointMA in all cases. This suggests that model selection approaches are generally less stable than model averaging methods, which is consistent with the findings of \citet{Yuan2005combining}.
\begin{table}[!ht]
	\centering

		\caption{FPE for the stock return data under rolling and recursive estimation windows}
		\label{tab.case1}
		\renewcommand{\arraystretch}{0.6}
		\small 
		\setlength{\tabcolsep}{18pt}
			\begin{threeparttable}
		\begin{tabular}{lllllll}
			\toprule
			$\tau$ & $T_1$ & ESMA & FM & Robust & HighDim & JointMA \\
			\midrule
			\multicolumn{7}{c}{Panel A: Rolling estimation window} \\
			\midrule
			0.05 & 100 & 0.270 & 0.810 & 0.781 & 0.418 & \textbf{0.250} \\
			0.05 & 200 & 0.229 & 0.394 & 0.405 & 0.331 & \textbf{0.228} \\
			0.05 & 400 & \textbf{0.175} & 0.226 & 0.215 & 0.255 & 0.219 \\
			0.1  & 100 & \textbf{0.209} & 0.427 & 0.483 & 0.469 & 0.216 \\
			0.1  & 200 & \textbf{0.171} & 0.271 & 0.272 & 0.378 & 0.194 \\
			0.1  & 400 & \textbf{0.166} & 0.196 & 0.183 & 0.313 & 0.205 \\
			\bottomrule
			\multicolumn{7}{c}{Panel B: Recursive estimation window} \\
			\midrule
			0.05 & 100 & \textbf{0.164} & 0.261 & 0.265 & 0.261 & 0.206 \\
			0.05 & 200 & \textbf{0.161} & 0.209 & 0.212 & 0.242 & 0.207 \\
			0.05 & 400 & \textbf{0.154} & 0.191 & 0.190 & 0.219 & 0.218 \\
			0.1  & 100 & \textbf{0.136} & 0.171 & 0.167 & 0.306 & 0.184 \\
			0.1  & 200 & \textbf{0.135} & 0.153 & 0.153 & 0.288 & 0.189 \\
			0.1  & 400 & \textbf{0.138} & 0.149 & 0.148 & 0.268 & 0.200 \\
			\bottomrule
		\end{tabular}
		\begin{tablenotes}\footnotesize
			\item[] Note:  Boldface indicates the best-performing method.
		\end{tablenotes}
	\end{threeparttable}
\end{table}

The ordering of nested candidate models can affect the performance of model averaging methods. To examine this issue, we consider an alternative strategy that ranks covariates by marginal quantile utility (MQU). The MQU is defined as $|Q_\tau(y | x_j) - Q_\tau(y)|$, where $Q_\tau(y | x_j)$ denotes the $\tau$-th conditional quantile of $y$ given $x_j$, and $Q_\tau(y)$ denotes the unconditional $\tau$-th quantile. A larger MQU indicates a stronger dependence between $y$ and $x_j$ at the $\tau$-th quantile; see \citet{Wang2023} for details. Table~\ref{tab.case1_mqu} reports the forecasting results when candidate models are ordered according to the MQU criterion. The overall patterns are similar to those obtained using correlation-based ordering. In particular, recursive estimation further improves forecasting accuracy, and ESMA remains the most stable performer across all configurations.
\begin{table}[!ht]
	\centering
		\caption{FPE for the stock return data sorted by MQU under rolling and recursive estimation windows}
		\label{tab.case1_mqu}
		\renewcommand{\arraystretch}{0.6}
		\small 
		\setlength{\tabcolsep}{18pt}
		\begin{threeparttable}
		\begin{tabular}{lllllll}
			\toprule
			$\tau$ & $T_1$ & ESMA & FM & Robust & HighDim & JointMA \\
			\midrule
			\multicolumn{7}{c}{Panel A: Rolling estimation window} \\
			\midrule
			0.05 & 100 & 0.240 & 0.815 & 0.786 & 0.423 & \textbf{0.233} \\
			0.05 & 200 & 0.226 & 0.398 & 0.409 & 0.335 & \textbf{0.207} \\
			0.05 & 400 & \textbf{0.175} & 0.228 & 0.217 & 0.256 & 0.180 \\
			0.1  & 100 & 0.211 & 0.430 & 0.487 & 0.472 & \textbf{0.192} \\
			0.1  & 200 & \textbf{0.165} & 0.276 & 0.277 & 0.383 & 0.171 \\
			0.1  & 400 & \textbf{0.160} & 0.193 & 0.180 & 0.309 & 0.166 \\
			\bottomrule
			\multicolumn{7}{c}{Panel B: Recursive estimation window} \\
			\midrule
			0.05 & 100 & \textbf{0.171} & 0.265 & 0.269 & 0.266 & 0.175 \\
			0.05 & 200 & \textbf{0.167} & 0.213 & 0.217 & 0.247 & 0.174 \\
			0.05 & 400 & \textbf{0.152} & 0.194 & 0.193 & 0.222 & 0.163 \\
			0.1  & 100 & \textbf{0.140} & 0.172 & 0.169 & 0.308 & 0.147 \\
			0.1  & 200 & \textbf{0.136} & 0.155 & 0.155 & 0.289 & 0.148 \\
			0.1  & 400 & \textbf{0.134} & 0.146 & 0.145 & 0.265 & 0.150 \\
			\hline
		\end{tabular}
		\begin{tablenotes}\footnotesize
			\item[] Note:  Boldface indicates the best-performing method.
		\end{tablenotes}
	\end{threeparttable}
\end{table}

\subsection{U.S. GDP Growth}

In recent years, expected shortfall has been increasingly used to measure macroeconomic risks. A prominent example is \citet{adrian2019vulnerable}, who employ ES to assess downside risks in U.S. GDP growth. Using a regression-based framework, they link future ES of GDP growth to current macroeconomic and financial conditions, with GDP growth and the National Financial Conditions Index (NFCI) as key covariates.

Following this line of research, we apply the proposed ESMA method to forecast one-quarter-ahead U.S. GDP growth using two widely used indicators: the NFCI and the Chicago Fed National Activity Index (CFNAI).\endnote{Data source: 1. U.S. GDP: \url{https://fred.stlouisfed.org/series/GDP}\quad 2. NFCI: \url{https://www.chicagofed.org/research/data/nfci/current-data}\quad 3. CFNAI: \url{https://www.chicagofed.org/research/data/cfnai/current-data}} The NFCI is a weighted average of 105 measures of financial conditions, while the CFNAI aggregates 85 indicators of real economic activity. Since the NFCI is available at weekly frequency and the CFNAI at monthly frequency, we convert both series to quarterly frequency by averaging observations within each quarter, consistent with \citet{adrian2019vulnerable}.

Let $Y_t$ denote the quarterly year-on-year U.S. GDP growth rate. To capture the persistence and cyclical patterns in GDP growth, we include up to four lagged terms of $Y_t$ in addition to the NFCI and CFNAI. The full specification therefore contains seven covariates, including an intercept. We consider $2^6$ non-nested candidate models, each of which includes an intercept term.

The sample covers the period from the first quarter of 1974 to the second quarter of 2025, yielding 206 observations. We evaluate forecasting performance using both recursive and rolling estimation windows, with initial window sizes $T_1 = 48, 72,$ and $96$ quarters. Forecast accuracy is assessed using the same criterion as in Section~5.1.

Table~\ref{tab.gdp_rolling_recursive} reports the out-of-sample prediction errors under both rolling and recursive estimation windows. First, ESMA delivers stable and competitive forecasting performance across different quantile levels and window schemes. In particular, under the recursive window, ESMA consistently achieves the smallest prediction errors for both $\tau=0.05$ and $\tau=0.1$, indicating that the proposed method can effectively produce reliable ES forecasts. Second, under the rolling estimation window with $\tau=0.05$, especially when $T_1=48$ and $T_1=72$, the effective sample size is relatively small and, in some cases, even comparable to or smaller than the number of candidate models. In this setting, model averaging methods tend to be less stable, and the HighDim method performs particularly well and achieves the smallest FPE. Third, across both rolling and recursive windows, JointMA performs poorly with prediction errors substantially larger than those of other methods, especially under the recursive window. A possible explanation is that JointMA involves optimizing over $2^7$ parameters simultaneously, which can lead to unstable estimation in small samples. In contrast, the proposed two-stage ESMA separates the estimation of VaR and ES, leading to more stable estimation and better finite-sample performance. The GDP results therefore highlight the practical advantage of the two-stage approach when forecasting macroeconomic downside risks.

\begin{table}[!ht]
	\centering
		\caption{FPE for the GDP data under rolling and recursive estimation windows}
		\label{tab.gdp_rolling_recursive}
		\renewcommand{\arraystretch}{0.6}
		\small 
		\setlength{\tabcolsep}{18pt}
		\begin{threeparttable}
		\begin{tabular}{lllllll}
			\toprule
			$\tau$ & $T_1$ & ESMA & FM & Robust & HighDim & JointMA \\
			\midrule
			\multicolumn{7}{c}{Panel A: Rolling estimation window} \\
			\midrule
			0.05 & 48 & 0.149 & 0.266 & 0.266 & \textbf{0.095} & 0.411 \\
			0.05 & 72 & 0.210 & 0.466 & 0.683 & \textbf{0.078} & 1.956 \\
			0.05 & 96 & 0.210 & 0.139 & 0.151 & \textbf{0.140} & 0.363 \\
			0.1  & 48 & \textbf{0.052} & 0.161 & 0.101 & 0.078 & 0.280 \\
			0.1  & 72 & \textbf{0.074} & 0.077 & 0.126 & 0.103 & 0.329 \\
			0.1  & 96 & \textbf{0.054} & 0.061 & 0.055 & 0.079 & 0.242 \\
			\bottomrule
			\multicolumn{7}{c}{Panel B: Recursive estimation window} \\
			\midrule
			0.05 & 48 & \textbf{0.106} & 0.353 & 0.138 & 0.140 & 2.331 \\
			0.05 & 72 & \textbf{0.119} & 0.413 & 0.159 & 0.157 & 2.609 \\
			0.05 & 96 & \textbf{0.127} & 0.477 & 0.167 & 0.181 & 3.125 \\
			0.1  & 48 & \textbf{0.064} & 0.092 & 0.074 & 0.091 & 0.411 \\
			0.1  & 72 & \textbf{0.070} & 0.100 & 0.077 & 0.103 & 0.377 \\
			0.1  & 96 & \textbf{0.073} & 0.101 & 0.074 & 0.115 & 0.392 \\
			\bottomrule
		\end{tabular}
		\begin{tablenotes}\footnotesize
			\item[] Note:  Boldface indicates the best-performing method.
		\end{tablenotes}
	\end{threeparttable}
\end{table}

\section{Conclusion} \label{chap:conclusion}
This paper proposes a two-stage cross-validation model averaging procedure for predicting expected shortfall. The first stage applies quantile model averaging, while the second stage adopts an MSE-based model averaging approach. From a theoretical perspective, when the set of candidate models contains correct models, the estimated coefficients are consistent and the sum of the model averaging weights in the correct models converges to one. When all candidate models are misspecified, the proposed procedure remains asymptotically optimal. Simulation studies and empirical applications further demonstrate the superior performance of the proposed method.

Several extensions are worth further investigation. First, financial and economic systems often operate in changing environments and involve high-dimensional predictors. This motivates the study of expected shortfall prediction using two-stage time-varying factor-augmented model averaging frameworks; see, for example, \citet{Chen2024Time-varying}. Second, the relationship between expected shortfall and predictors may be highly nonlinear and could be modeled using flexible machine learning methods, such as neural networks. Extending the proposed model averaging procedure to more general and nonparametric models is an interesting direction for future research.

\theendnotes

 \bibliography{MA_PLM_ref.bib}

\end{document}